# Phonon-assisted Auger enables ultrafast charge transfer in CdSe Quantum Dot/Organic Molecule

Zhi Wang[1,2], Mona Rafipoor[1,2], Pablo García Risueño[1,2], Jan-Philip Merkl[1], Peng Han[1,3], Holger Lange[1,2] and Gabriel Bester[1,2*]

[1] Institut für Physikalische Chemie, Universität Hamburg, Grindelallee 117, D-20146 Hamburg, Germany.

[2] The Hamburg Centre for Ultrafast Imaging, Luruper Chaussee 149, D-22761 Hamburg, Germany

[3] Department of Physics, Capital Normal University, Beijing Key Lab for Metamaterials and Devices, Beijing 100048, China

*gabriel.bester@uni-hamburg.de

## ABSTRACT

Charge transfer between photoexcited quantum dots and molecular acceptors is one of the key limiting processes in most applications of colloidal nanostructures, most prominently in photovoltaics. An atomistic detailed description of this process would open new ways to optimize existing and create new structures with targeted properties. We achieve a one-to-one comparison between *ab-initio* non-adiabatic molecular dynamics calculations and transient absorption spectroscopy experiments, which allows us to draw a comprehensive atomistic picture of the charge transfer process, following the time evolution of the charge carrier across the electronic landscape and identifying the thereby induced vibrations. For two quantum dot sizes we find two qualitatively different processes. For the larger structure we find a relatively slow ($\tau = 516$ fs) transfer process that we explain by the existence of a large energy detuning and weak vibronic coupling. For the smaller structure the process is ultrafast ($\tau = 20$ fs) due to an efficient, phonon-assisted Auger process triggered by a strong electron-hole coupling.

# INTRODUCTION

The coupling of nanostructures to the environment and their ability to transfer charge or energy defines their usability for nearly all possible applications. For instance, photovoltaic or catalytic devices require an efficient charge separation following the optical transition. Similarly, in biomedical imaging and display technology good optical properties need to be combined with specific charge transfer mechanisms. These ultrafast charge transfer (CT) processes are governed by a non-trivial combination of the solvent, the energetic alignment between the excited states, the overlap between the orbitals of the donor-acceptor pair, and the coupling between electronic degrees of freedom and the nuclear motion, called vibronic coupling[1-3].

Experimentally, ultrafast pump-probe techniques can deliver valuable information in the appropriate femtosecond time domain and quantify the reaction rates. The electronic excitation is thereby created by an ultrashort optical pulse followed, with varying time delay, by a probe pulse delivering information on the time-dependent occupation of certain states. Theoretically, a mesoscopic description for the transfer rate in CT processes was developed earlier by Marcus[4] using concepts such as the reorganisation energy of the surrounding and an exponential dependence on activation energies. The remaining big open questions revolve around the interplay between the vibrational degrees of freedom, which are eventually absorbing the excess energy of the electronic system, and the time evolution of the electronic states. In other words, the full description of the CT path, unravelling the ``quantum molecular movie'' of coupled electronic and ionic degrees of freedom. Recently, real-time time-dependent density functional theory (rt-TDDFT) has evolved as a powerful tool to study electron dynamics, as demonstrated in recent publications concerning, e.g., plasmons[5], the kinetic ion collision[6], as well as in the context of photovoltaics[7-10]. This approach allows, to a certain degree, to follow the relaxation and transfer of the excited particle across the landscape of electronic states. Although very powerful, most of the rt-TDDFT calculations have been restricted to extremely short time periods and small systems thus lacking experimental validation.

In this work, by combining a high-efficient rt-TDDFT non-adiabatic molecular dynamics (NAMD) method[6] with ultrafast white-light transient absorption (TA) spectroscopy, both performed on the structurally same, specifically designed QD-MV$^{2+}$ CT system, we are able to follow the CT path. As prerequisite, we validate the quality of our theoretical approach by an unprecedented "one-to-one" comparison and agreement between the calculation and the experiment, especially for the onset of the MV$^+$ signal (the indication for the completion of the CT process). We find two qualitatively different CT process, i.e. two different paths, for different

QD sizes: for our large QD (diameter ≈ 1.7 nm), the QD-MV$^{2+}$ CT rate ($\tau = 516$ fs) mostly depends on the intramolecular carrier cooling, which is dominated by vibronic coupling; for a smaller QD with diameter of 1.2 nm, a stronger electron-hole coupling exists, which allows for an efficient phonon-assisted Auger transfer, leading to a much higher CT rate ($\tau = 20$ fs). Finally, we identify which vibrations are excited during and shortly after the phonon-assisted Auger CT. Such a detailed understanding will be helpful to design nano-architectures with specific properties and eventually to dynamically steer CT processes by exciting specific vibrations.

## RESULTS AND DISCUSSION

**Experimental TA spectra.** Figure 1 (a) and (b) display TA maps of the 1.8 nm CdSe QD with and without acceptor MV$^{2+}$ for the first 2 ns after excitation in the wavelength regime of 425-725 nm. The two maps differ significantly. While the plain QD signal is almost constant in the displayed time windows, the combined system features a rich dynamics. Figure 1 (c) shows the early state of dynamics of the combined system. Figure 1 (d) and (e) are showing the dynamics directly after the signal rises (220 fs) and at a later time (700 fs). Initially, the two spectra are very similar. They consist of the QD exciton bleach centered around 460 nm and a photo-induced absorption around 500 nm[11]. The exciton lifetime is in the 20-ns regime, thus the plain QD spectra have not evolved during the first ps, however in the QD-MV$^{2+}$ sample the electron transfer from QD to MV$^{2+}$ results in a faster recovery of the QD exciton bleach and in the appearance of a positive absorption change at longer wavelengths (550-700 nm), which can be related to the change from MV$^{2+}$ to MV$^+$ [12-14]. These two peaks allow us a discussion of the dynamics of the CT process: we have access to the QD carrier concentration via the QD bleach and we can probe the additional electron on the MV$^{2+}$ via the rise of the MV$^+$ absorption.

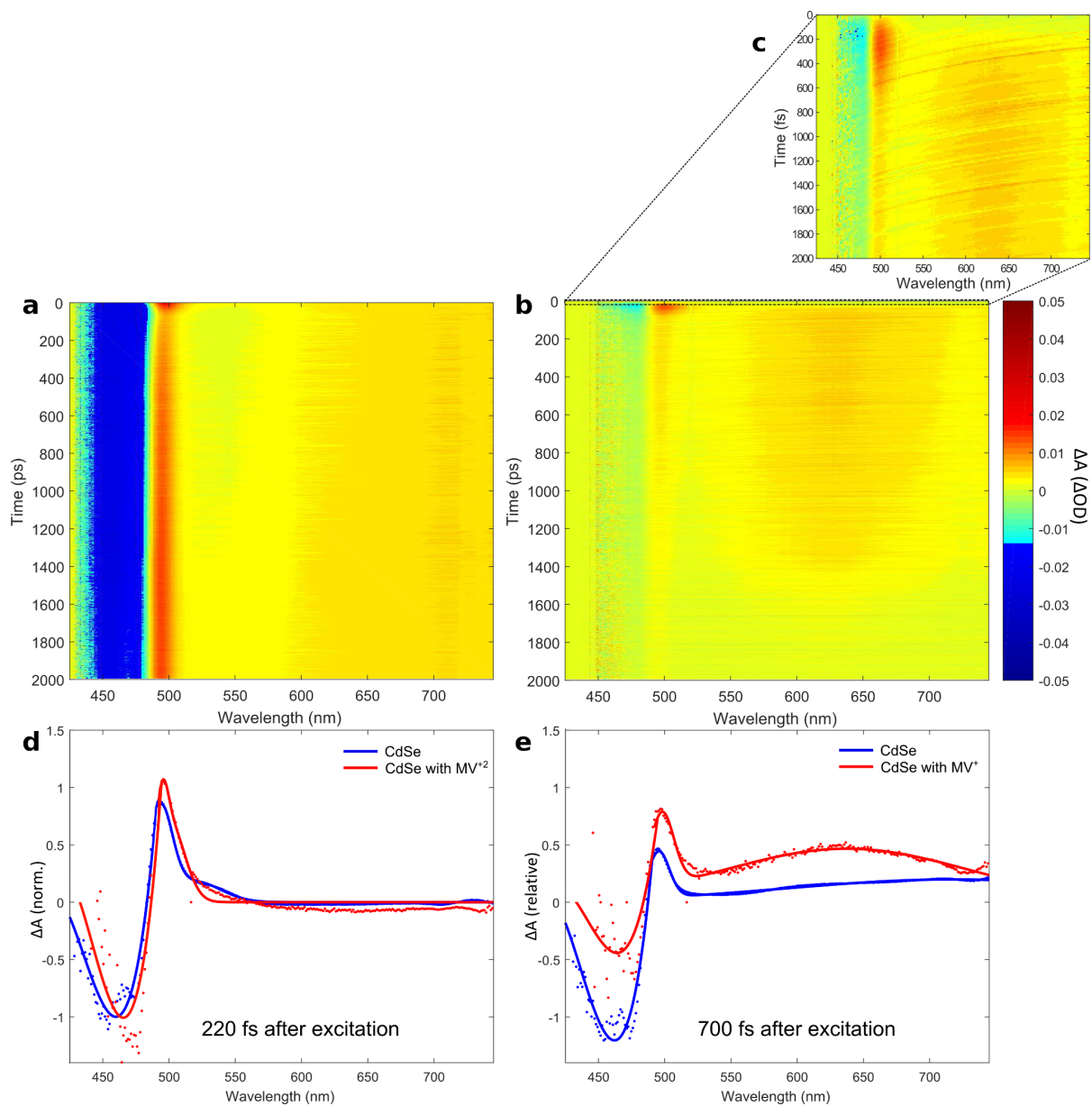

**Figure 1 | TA spectra.** (a) CdSe QD only and (b) CdSe QD/MV$^{2+}$ TA maps for excitation at 400 nm. (c) CdSe QD/MV$^{2+}$ TA map for the first 2 ps. (d) TA spectra for CdSe QD (blue) and CdSe QD/MV$^{2+}$ (red) at 220 fs after excitation. (e) same as (d) but for 700 fs after excitation. The spectra are normalized to the QD absorption bleach intensity at 220 fs after photoexcitation (bleach maximum).

**Comparison of TDDFT-NAMD with TA spectra.** Figure 2 (a) shows the time evolution of the electron population of the LUMO state of CdSe QD with 1.7 nm diameter obtained from TDDFT-NAMD simulation (blue lines) and the measured TA signal (red lines) with wavelength of 462 nm, which corresponds to the HOMO-LUMO absorption peak of this QD. Figure 2 (b) shows the time evolution of the electrons on the MV$^{2+}$ LUMO obtained from TDDFT-NAMD (blue lines) and the TA signal at 600 nm (red lines) corresponding to the reduced MV$^+$ absorption peak center.

The structures used in the calculations are shown in detail in the Methods section. We see that the TDDFT-NAMD results for both the bleached QD and the reduced MV[+] signals are in excellent agreement with the experiment. Furthermore, the time between the onset of the QD and the MV[+] signals also agrees well with the experiments. Specifically, after 500 fs both TDDFT-NAMD and TA show about 70% decrease of the QD signal, and a 230-fs delay is observed before the excited electron occupies the MV[2+] LUMO (MV[+] HOMO) in the TDDFT-NAMD calculation, compared with the ~200-fs delay between the bleached QD and reduced MV[+] signals in the TA spectrum.

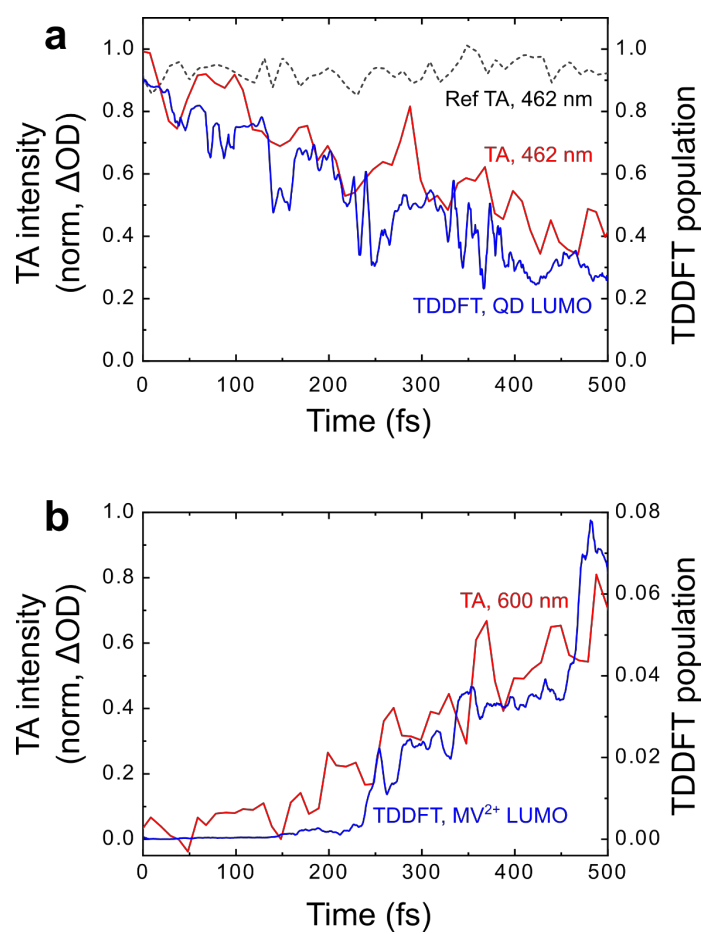

**Figure 2 | Comparison between experimental TA spectra and TDDFT-NAMD results.** (a) QD HOMO-LUMO absorption bleach recovery at 462 nm and (b) reduced MV+ absorption at 600 nm from TA (red lines), shown together with the TDDFT results (blue lines). Black dash line in (a) shows the reference QD absorption bleach without MV[2+] molecules. The 462 nm TA signal is normalized to its bleach maximum at 220 fs, while the 600 nm TA signal is normalized to its maximum at 3 ps.

**Size-dependent CT paths.** Reducing the size of the QD normally induces a larger donor-acceptor orbital overlap, a different band offset, and a stronger quantum confinement effect which results in a stronger electron-hole coupling. In Figure 3(b) we show for the first 400 fs the time evolution of the excited electron populations on the QD LUMO and the $MV^{2+}$ LUMO for a 1.2 nm and a 1.7 nm diameter QD with $MV^{2+}$ attached. We find a significantly more efficient CT in the smaller structure. Quantitatively, the exponential fit of the LUMO population in the 1.2 nm diameter QD yields $\tau = 20$ fs, compared to $\tau = 516$ fs in the 1.7 nm diameter QD. Accordingly, the $MV^{2+}$ LUMO population in the 1.2 nm diameter QD-$MV^{2+}$ rises after 116 fs, much earlier than the 230 fs in the larger QD system.

The band alignment of the two QD-$MV^{2+}$ systems are shown in Figure 3(a). The CT paths (time evolution of QD and molecular states) of the two systems are shown in Figure 3(c) and (d). We see that:

(1) In 1.7 nm diameter QD-$MV^{2+}$ the excited electron transfer primarily occurs between QD LUMO and $MV^{2+}$ LUMO+1, which is significantly faster than the intramolecular carrier cooling from $MV^{2+}$ LUMO+1 to $MV^{2+}$ LUMO. As a result, the relatively slow intramolecular carrier cooling represents the bottleneck of the whole CT process;

(2) In 1.2 nm diameter QD-$MV^{2+}$ there are 4 molecular states ($MV^{2+}$ LUMO to LUMO+3) in the band gap, thus a more complex carrier cooling route is anticipated. Moreover, the energy detuning between QD and molecular states are also larger than in the larger QD, which in principle should reduce the CT rate, in contradiction to the obtained results. Two major phases can be observed: the first phase, described by the QD LUMO occupation (blue curve in Figure 3(d)), lasts from 0 fs to ~70 fs, where the carrier transfers from the QD to the molecule, particularly from QD LUMO to $MV^{2+}$ LUMO+2 (green curve in Figure 3(d)); the overlapping second phase, starts as early as ~ 40 fs and extends to the end of the simulation time and follows the pathway: $MV^{2+}$ LUMO+2, to $MV^{2+}$ LUMO+1, finally to $MV^{2+}$ LUMO. The two phases are not explicitly separated in time but co-exist approximately between 40 fs and 70 fs.

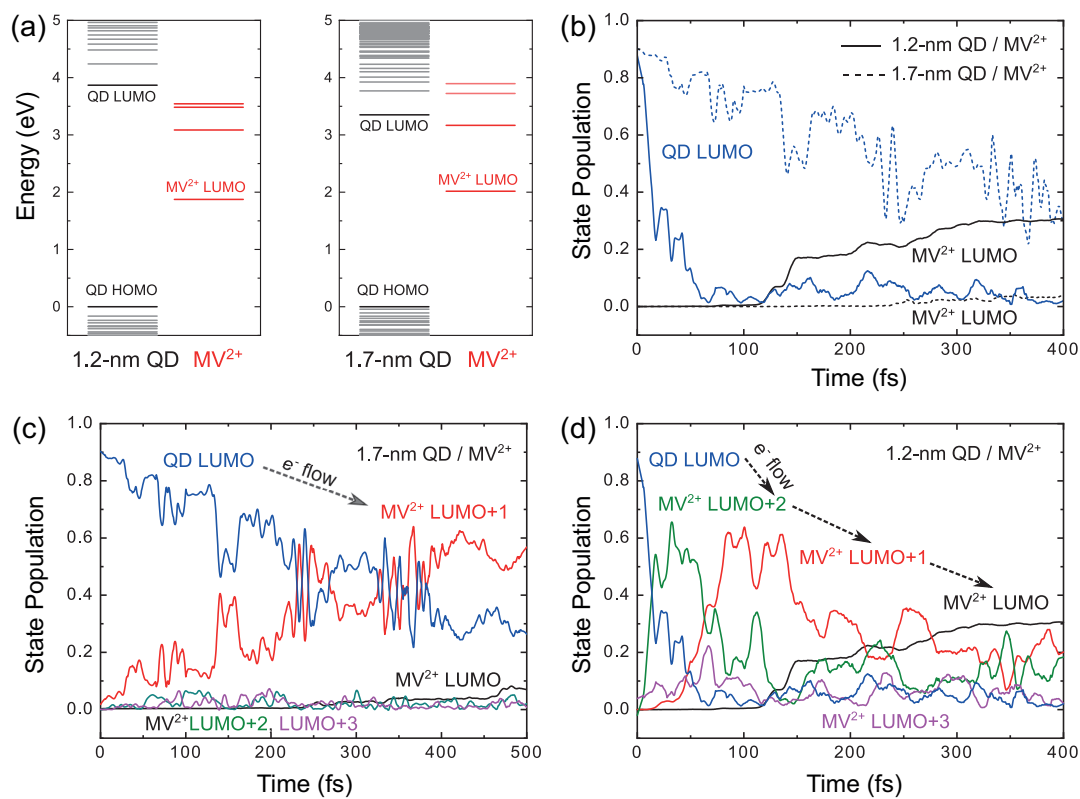

**Figure 3 | Charge transfer paths for small and large QDs.** (a) Band offset between QD and $MV^{2+}$ in 1.7-nm (large QD) and 1.2-nm (small QD) diameter QDs attached to $MV^{2+}$. (b) Electron populations on QD LUMO and $MV^{2+}$ LUMO states for the small QD (solid lines) and the large QD (dashed line). (c) Populations of the QD LUMO and molecular states $MV^{2+}$ LUMO to $MV^{2+}$ LUMO+3 in the large QD. (c) Same as (b) in the small QD. All results come from TDDFT-NAMD calculations.

**Phonon-assisted Auger CT process.** While the importance of the ion dynamics is clear in the 1.7 nm diameter QD-$MV^{2+}$ where the relatively slow intramolecular carrier cooling experiences a mild bottleneck, it is questionable for the smaller QD-$MV^{2+}$ system. Indeed, the ultrafast charge transfer of $\tau = 20$ fs in the first phase of dynamics could be regarded as too fast to involve any effective couplings between electron and nuclear motion. To assess the importance of the nuclear vibration in the first CT phase, in Figure 4 (a) and (b) we show two levels of TDDFT calculations: level 1 is TDDFT-NAMD which contains all nuclear vibrational effects and level 2 is a frozen-ion TDDFT calculation where all ions are fixed during the entire simulation. Calculation details can be found in the Methods section. We find that:

(1) In the 1.2 nm diameter QD-$MV^{2+}$ frozen-ion TDDFT in Fig. 4(b), we see a significant Rabi-like oscillation on the population of the QD LUMO, which has a minimum value of 0.5 $e^-$ (electron charge) and an average value of 0.7 $e^-$. This energy-conserving large oscillation can be attributed to the strong QD-molecule wavefunction overlap, and we suggest that it is the signature of a coherent QD-$MV^{2+}$ state, where the charge density moves forth and back between QD and molecule but the electronic relaxation remains impossible due to the lack of vibrations (frozen-ion). This is in contradiction to the "simple" Auger-assisted electron transfer model[9, 15, 16] which describes a scattering process where phonons usually do not participate (they do, however, in the subsequent carrier cooling process). The results including vibronic coupling (solid blue line) are entirely different and show a very fast depopulation of the QD LUMO which hints at an Auger-type process involving vibrations: a phono-assisted Auger CT. The involvement of phonons in our CT process is reasonable from a classical analogue of momentum conservation. The transferring electron transfers momentum to the vibration. We analyze the situation further in Figure 4(c) and (d), where we show the time-evolution of the energies of the excited electron and of the hole. We see that in Figure 4(d) during the initial phase for the small QD, from 0 to 100 fs, the electron and hole state both loose energy in a concerted fashion, which is the Auger signature: the electron cools down, by exciting the hole. After that ultrafast process, which leaves the QD LUMO state depopulated (see Fig.4(b)), the QD and the molecule are submitted to a low frequency (about 70 cm$^{-1}$) vibration of their spatial separation, as shown in Fig.4.(e). The quasi immediate transfer of the electron from QD to molecule generates an electric field which drives the QD-MV system in "mesoscopic" oscillations of both entities (the molecule readjusts its equilibrium distance from the QD harmonically). This effect gives rise to the long oscillation between 100-400 fs in the electron and hole energies.

(2) For the larger 1.7 nm diameter QD-MV$^{2+}$ system, if all nuclei are frozen, the CT of the excited electron is entirely absent in Fig.4(a) (dashed line), and only very small Rabi-like oscillations can be observed. It means that the population transfer observed in the non-adiabatic case (solid line) is a consequence of the vibronic coupling, missing of which will result in the inability of the electronic degrees of freedom to release energy to the ionic motion, thus blocking the most important carrier cooling route. The electron and hole energies are shown in Fig.4(c) and show a moderate decrease of energy, with a growing separation of the electron from the hole energy curves. We attribute this to the loss of exciton binding energy. The electron transfers "slowly" away from the QD (Fig.4(a)) and thereby loses exciton binding, increasing the eigenvalue of the electron and decreasing the one of the hole. We see no clear signature of Auger in this process and the time scale corresponds to the standard electron cooling process dominated by phonon-lifetime[19] ,in the ps range.

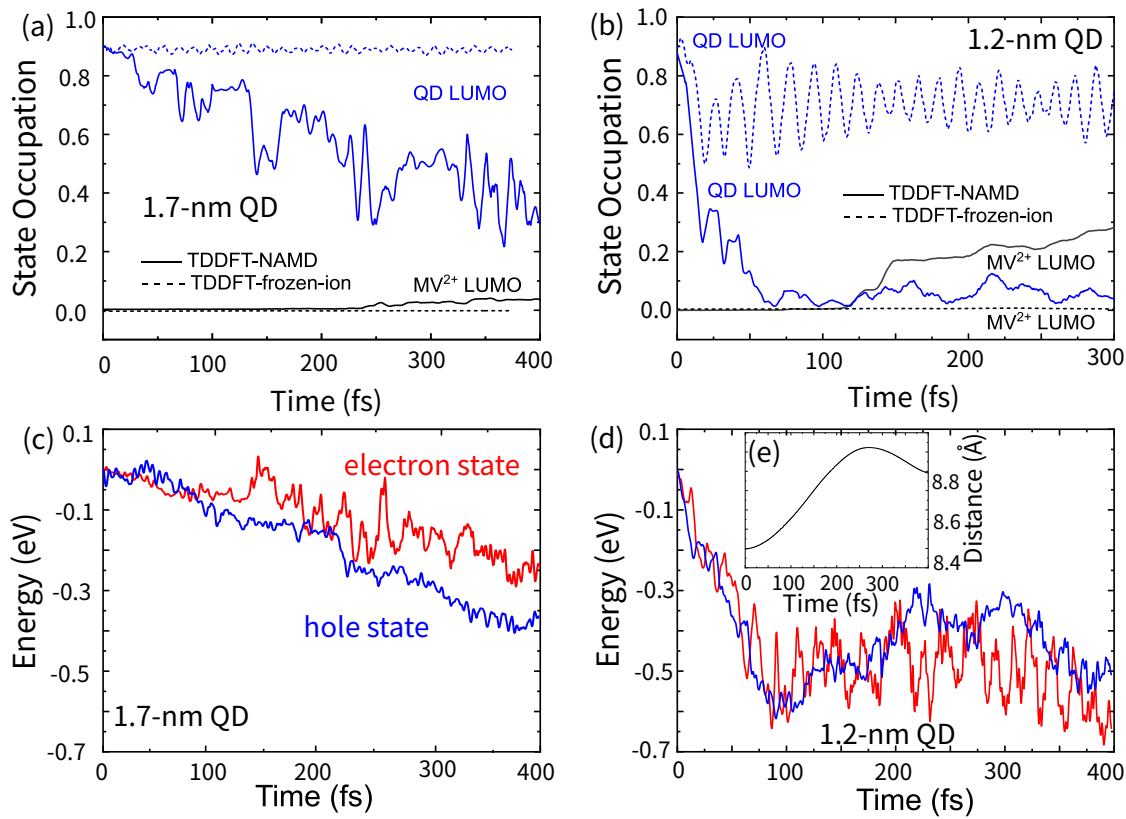

**Figure 4 | Evidence for the phonon-assisted Auger CT.** (a) Populations of the QD LUMO and $MV^{2+}$ LUMO from TDDFT-NAMD (solid lines) and frozen-ion TDDFT (dash lines) for the large QD/$MV^{2+}$ system; (b) same properties as (a) for the small QD/$MV^{2+}$ system. (c) TDDFT-NAMD excited electron and hole energies in the large QD/$MV^{2+}$ system; (d) same as (c) but in the small QD/$MV^{2+}$ system; (e) average distance between the QD and the $MV^{2+}$ molecule.

**Analysis of the electron-nuclear vibrational coupling.** In TDDFT-NAMD, the nuclei are driven by the ab-initio forces obtained from the time-dependent (TD) charge density, which means that the coupling between electronic and ion dynamics is naturally included. This coupling contains roughly two components: the couplings of nuclear motion with (a) *the excited electron*, and (b) *all other none-excited electrons*. The comparison between TDDFT-NAMD and frozen-ion TDDFT shown in the last paragraph yields the total coupling of (a) and (b), and the comparison between TDDFT-NAMD and BOMD, where the nuclei are still driven by the ab-initio forces but the electronic system is always in the ground state, will reveal coupling (a) only.

To reduce the highly complex and multidimensional coordinate space to an analysis in terms of key modes[17], we expand the TD nuclear trajectories $\{R_\kappa^{NAMD}(t)\}$, $\{R_\kappa^{BOMD}(t)\}$ ($\kappa$ is the atom index, $\kappa = 1, ..., N$) from TDDFT-NAMD and BOMD, respectively, on the basis set of the eigendisplacements $\{U_{\kappa\alpha}^\nu\}$ where $U_{\kappa\alpha}^\nu = \xi_{\kappa\alpha}^\nu/\sqrt{M_\kappa}$ ($\alpha$ is the Cartesian coordinate indices) from the adiabatic harmonic modes $\nu$ ($1 \leq$ $\nu \leq 3N - 6$) with eigenfrequencies $\{\omega_\nu\}$.[18] The expansion coefficients are denoted as $\{a_\nu^{NAMD}(t)\}$, $\{a_\nu^{BOMD}(t)\}$, and their Fourier transform as $\{a_\nu^{NAMD}(\omega)\}$, $\{a_\nu^{BOMD}(\omega)\}$. The calculations necessary to obtain $\{U_{\kappa\alpha}^\nu\}$ and $\{a_\nu(\omega)\}$ are given in the Methods section. Note that if the electronic system stays in the ground state and all atoms vibrate harmonically close to their ideal nuclear positions, $a_\nu(\omega)$ will have non-zero value only if $\omega = \omega_\nu$. We extract the vibrational energy redistribution from the difference in vibrational mode occupations $a_\nu(\omega)$, i.e., $\Delta a_\nu(\omega) \equiv a_\nu^{NAMD}(\omega) - a_\nu^{BOMD}(\omega)$. Alternatively, we calculate the coupling matrix elements $g^\nu(i,j)$ where $i, j$ = MV$^{2+}$ LUMO to LUMO+3 explicitly. We present $\Delta a_\nu(\omega)$ as bubbles in Figure 5(a) and $|g^\nu(i,j)|$ as bars in Figure 5(b). The diagonal terms $g^\nu(i,i)$ describe the coupling of the electronic state $i$ to the vibration $\nu$, while the off-diagonal terms $g^\nu(i,j)$ describe the transition from state $i$ to state $j$ enabled by the vibration $\nu$. We find that:

(1) Some modes experience mainly a frequency red-shift, which can be seen in Figure 5(a) as two circles with opposite color in vertical proximity (e.g. modes 12, 25, 26, 30, 32, 37, 53, 55, 60, 70). This red-shift is compatible with the simple chemical picture that the excitation has loosened the bonds via the occupation of antibonding orbitals and therefore softened the modes.

(2) Some modes are genuinely excited by the electron dynamics. Most prominently, the modes 17, 45, 47, 48, 54, 61, 62 and 63 show the largest contribution to the CT process as indicated by the largest $|\Delta a_\nu(\omega)|$ values in Figure 5(a). Mode 17 is an out-of-plane (i.e. towards the QD) torsion of the pyridyl ring, modes 61-63 are out-of-plane C-H wagging, mode 47 is C-H stretching and bending on the CH$_3$ ligand, and modes 45, 48 and 54 consist of in-plane C-H bending, C-C bending and stretching and C-H stretching and bending on the CH$_3$ ligands. The contributions of ring breathing, beating and ring C-H stretching are negligible. We show these prominent modes in Table S3 and movies in Supplementary Information. Among all prominent modes, the modes 17, 48, 54, 61 and 62 correspond to modes that effectively couple the electronic states MV$^{2+}$ LUMO, LUMO+1, LUMO+2 and LUMO+3, as indicated by the large off-diagonal matrix elements in Figure 5(b). The remaining dominant modes 45, 47 and 63 correspond to modes with vanishing off-site but large on-site $|g^\nu(i,i)|$ matrix elements. These modes are strongly coupled to the electronic states MV$^{2+}$ LUMO to LUMO+3. After the excitation, in form of the electron-hole pair creation, these electronic states experience a new potential energy surface (PES). For the strongly coupled modes, the excited PES is shifted in nuclear coordinate space and the vertical instantaneous excitation creates vibrational modes, just as expected from the Franck-Condon picture. On the other hand, the weakly coupled modes result in unshifted PES thus no vibrations are created.

(3) Some vibrational modes with rather large matrix elements $|g^v(i,j)|$, e.g. modes 33, 39, 60 and 74, have a very weak $\Delta a_v(\omega)$. We can find four main reasons for this seeming contradiction. First, the phonon-assisted Auger process we put forward in this work requires phonons that are not necessarily the ones most strongly coupled to the electronic states. Second, the $g^v(i,j)$ are calculated for the isolated molecule and hence neglect effects of the QD-molecule coupling. Third, in the $g^v(i,j)$ calculations we use the harmonic approximation by construction, while in TDDFT-NAMD anharmonic terms are naturally included. Finally, $g^v(i,j)$ are calculated for the ground-state, which means that the empty molecular states LUMO to LUMO+3 do not correspond to the TDDFT-NAMD which include excitation.

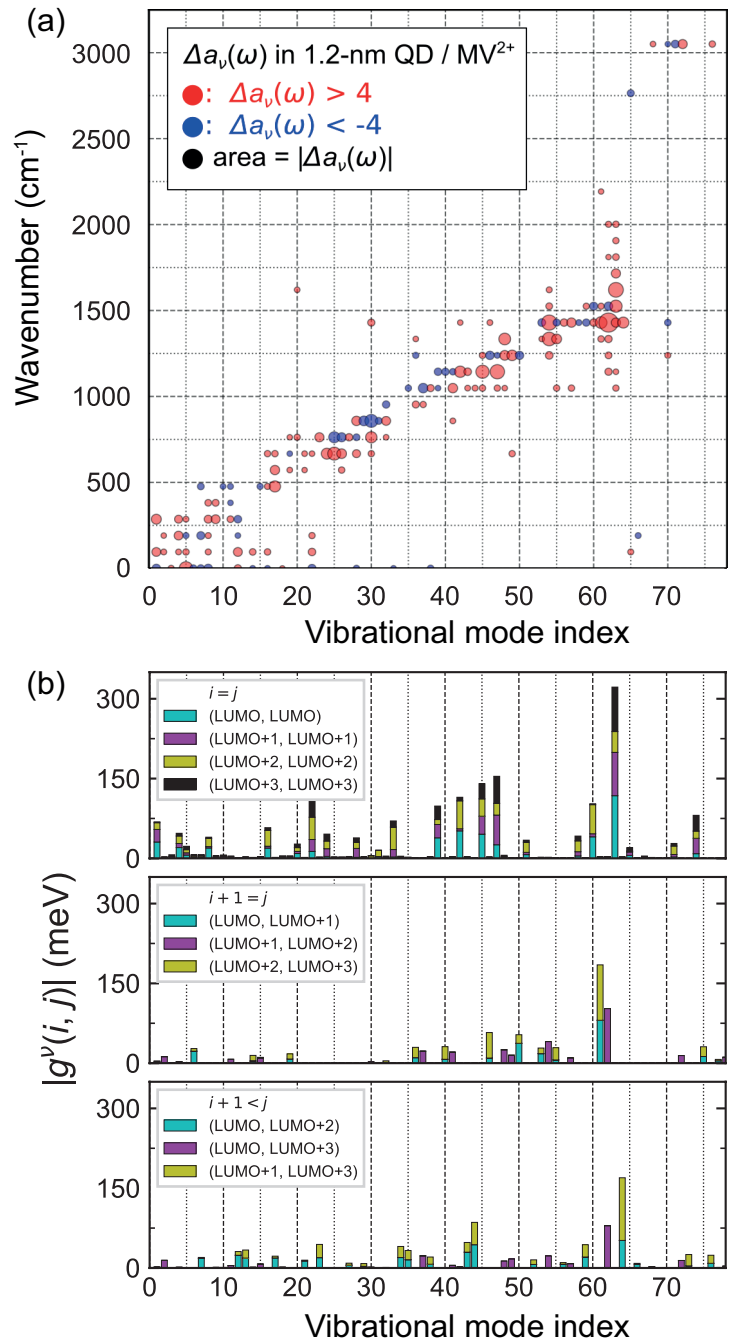

**Figure 5 | Analysis of the vibrational modes excited during and directly after CT.** (a) Bubble chart of the vibrational energy redistribution $\Delta a_v(\omega) \equiv a_v^{TD}(\omega) - a_v^{BO}(\omega)$ extracted from the first 400 fs in the small QD system. The area of the bubbles is proportional to the absolute value of $\Delta a_v(\omega)$, while the red and blue colors mean $\Delta a_v(\omega) > 0$ and $\Delta a_v(\omega) < 0$, respectively. (b) Electron-phonon coupling matrix element $|g^v(i,j)|$ for $i, j$ = MV$^{2+}$ LUMO to LUMO+3. The upper, middle and lower subplots in (b) show the cases when $i = j$, $i + 1 = j$, and $i + 1 < j$, respectively.

## CONCLUSIONS

We have achieved a one-to-one comparison between large scale *ab-initio* TDDFT-NAMD calculations and ultrafast white-light TA spectroscopy for the CT process in CdSe QD-MV$^{2+}$ system and obtain very good agreement, validating the numerical results. We use two sizes of QDs that reveal qualitatively two different types of CT and carrier relaxation processes.

For the large QD (d = ~1.7 nm), the alignment of the QD LUMO and the molecular LUMO+1 levels should be favorable to the CT process, however, the transfer remains relatively slow ($\tau$ = 516 fs) due to a large energy detuning between the vibrational modes and the molecular electronic gap. Vibronic coupling is too weak to allow for sub-ps relaxation[19] across this energy gap.

For the small QD (d~1.2 nm) we have four molecular levels within the QD bandgap and a less favorable band alignment situation. Nevertheless, the CT process ($\tau$ = 20 fs) and subsequent carrier relaxation is ultrafast, in the 100 fs range. We show that the small size of the donor-acceptor pair leads to a large electron-hole Coulomb coupling which enables an efficient phonon-assisted Auger CT process, where the hole residing on the QD is excited by the electron climbing down the ladder of electronic states on the molecule. We further show which vibrations are excited on the molecule during and immediately after the CT process and which are presumably involved in the phonon-assisted Auger process. This type of phonon-assisted Auger process should be generally possible on nanostructure/molecular systems as long as their coupling is sufficiently strong (i.e. for small QDs). For molecular redox systems the likelihood of the phonon-assisted Auger will depend, in addition to the required strong coupling, on the availability of resonant hole states. A full understanding of these process may open the way to new structure design, or even offer opportunities to drive CT process dynamically, by exciting specific vibrations.

# Methods:

**I. Chemicals**. 1-octadecene, hexadecylamine, methylviologene and oleic acid (OA) were purchased from Aldrich, trioctylphosphine (TOP) from abcr GmbH, cadmium oxide (99.998%) from Alfa Aesar, Selenium (99.999%) from chemPUR GmbH, toluene from Fischer Chemicals and 2-propanol, *n*-hexane and methanol from VWR Chemicals. All chemicals were used as delivered.

**II. QD-MV synthesis**. The QDs were prepared using a slightly modified protocol from Hens and coworkers[20]. Cadmium oxide was mixed with oleic acid in a Cd:OA ratio of 1:3. The mixture was degassed for 1 h at 100 °C under nitrogen flow and then heated up to 250 °C. During this process vacuum was applied (3x approx. 1 min) to remove the formed moisture. TOP was mixed with selenium and heated up to 200 °C to produce TOPSe (c(TOPSe)=1 M). Cadmium oleate (0.2 mmol with respect to Cd(II)) and hexadecylamine (0.6 mmol) were mixed with 10 mL 1-octadecene and degassed for 1 h at 100 °C under nitrogen flow. The reaction mixture was heated up to 180 °C and TOPSe (2 mmol) was injected rapidly. After 40 s 10 mL *n*-hexane and subsequently 10 mL toluene were added to cool the reaction mixture. After cooling to room temperature, the solvents were removed in vacuum and the particles were precipitated twice by adding approx. 10 mL of a 1:1 mixture of 2-propanol and methanol and stored in degassed (4 cycles of freeze-pump-thaw) chloroform. Methanol was degassed using 4 cycles of freeze-pump-thaw and a stock solution of methylviologene in degassed methanol was prepared. The QDs were mixed with methylviologene under nitrogen atmosphere to achieve a molar ratio of approx. 1QD:1MV$^{2+}$ in solution. The concentration of CdSe was determined from the absorption spectra using the equation of the *Mulvaney* group[21]. The size of QD was measured by TEM image analysis using a JEOL JEM-1011 microscope. See Supporting Information section II for details.

**III. TA measurements**. Pump-probe TA spectroscopy was performed using a commercial TA setup (Helios; Ultrafast Systems). A commercial amplified Ti-sapphire laser system (Spitfire-Ace, 800 nm, 6 W, 1 kHz, 35 fs; Spectra Physics) was employed to generate pump and probe pulses. The pump pulses were generated in an optical parametric amplifier (TOPAS-Prime; Light Conversion) with frequency mixer (NirUVis; Light Conversion) and chopped at 500 Hz. The probe beam was a broadband continuum white-light with a spectral range of 420-750 nm. The instrument response in this wavelength regime was estimated to be below 200 fs. In the calculation, the initial condition of the system is a single HOMO-LUMO excitation. In order to achieve best comparability between experiments and calculations and to avoid multiexciton effects, we performed a close to bandgap, low power excitation (400 nm, 127 μJ-cm²) for all measurements. We refrained from full resonant excitation as the necessary deconvolution of pump and probe would reduce the time resolution. The average carrier population per QD after excitation was estimated to below 0.15. The thermalization takes approximately 220 fs after which the system is in a condition that resembles the calculations best. Hence, we selected 220 fs as the time zero of the TA spectra.

**IV. Atomistic models.** We have theoretically considered two CdSe QDs: one is structurally the same as our experimental sample with a diameter of 1.7 nm [(CdSe)$_{65}$], while the other one is smaller [(CdSe)$_{33}$] and corresponds to the 'magic number' cluster widely used in *ab initio* calculations[22, 23]. Although trioctylphosphine oxide (TOPO) is used to passivate the surface dangling bonds in the synthesis of the CdSe QD, time-dependent *ab initio* calculations for a fully TOPO-coated QD are computationally out of reach, thus we use pseudo-hydrogens with fractional charges as passivants[24]. Another crucial element relates to the attachment of the MV$^{2+}$ complex to the QD. A direct experimental determination has not been reported yet. Nevertheless, some earlier reports are valuable: 1) viologens prefer mostly to adsorb flat on metal and semiconductor surfaces[25, 26]; 2) the photoexcited electron transfers directly from the semiconductor surface to the viologen bipyridyl rings, rather than through its alkyl chains[27]; 3) the CT efficiency is proportional to the number of CT-active molecules binding to one QD[28]. Accordingly, we build the 1:1 QD-MV$^{2+}$ structure for two sizes of QDs. Specifically, on each QD one pair of hydrogens connected to Selenium has been removed to make space for a flat MV molecule; moreover, considering that MV$^{2+}$ is introduced in the form of (MV)Cl$_2$ in chloroform, we have replaced two Cd-connected hydrogens with 2 Cl atoms. This configuration naturally keeps the surface pseudo-hydrogens paired, the total system charge neutral[26] and the MV complex in the cation MV$^{2+}$ state. The structures of the two QD-MV$^{2+}$ systems have been optimized using the Vienna *Ab-initio* Simulation Package (VASP)[29] and are shown in Figure S1 (a) and (b).

**V. Corrections to DFT energy levels.** In the case of the electronic structure in CdSe and in the molecule, LDA suffers from its well-known underestimation of band gaps. In addition, the QD, the molecule and the non-vacuum solvent environment form a non-uniform dielectric function distribution in real space that will also affect electronic states. To tackle this issue, we have applied corrections on LDA energy levels as successfully done previously[30, 31]. For the QD-MV-solvent system at each time step $t$, firstly we decompose the TD-LDA single-particle wavefunctions $\varphi_i(r,t)$ into two parts using a real space mask function $m(r,t)$,

$$\varphi_i = \varphi_i m + \varphi_i(1-m) = \varphi_i^{QD} + \varphi_i^{MV}, \tag{1}$$

where

$$m(r,t) = \begin{cases} 1, & if\ r \in QD \\ 0, & if\ r \in MV \end{cases} \tag{2}$$

Secondly, the many-body corrections on the QD and the molecular conduction band (CB) and valence band (VB) states can be written as,

$$\Delta_{c,v}^{QD}(t) = \Sigma_{e,h}(QD{:}QD) + \Sigma_{e,h}(QD{:}MV^{2+}) + \Sigma_{e,h}(QD{:}sol), \tag{3}$$

$$\Delta_{c,v}^{MV}(t) = \Sigma_{e,h}(MV^{2+}{:}QD) + \Sigma_{e,h}(MV^{2+}{:}MV^{2+}) + \Sigma_{e,h}(MV^{2+}{:}sol), \tag{4}$$

$\Sigma_{e,h}(a{:}a)$ is the self-energy correction in compound $a$ (QD or $MV^{2+}$) for unoccupied and occupied state, respectively, and $\Sigma_{e,h}(a{:}b)$ is the polarization energy correction in compound $a$ that is caused by the dielectric mismatch between compounds $a$ and $b$. Finally, the new Hamiltonian $H'$ after correction can be written as,

$$H'(t) = \sum_i \sum_{\substack{a,b= \\ QD,MV}} \left(\varepsilon_i^{LDA}(t) + \Delta_i^{a,b}(t)\right) |\varphi_i^a(t)\rangle\langle\varphi_i^b(t)|, \tag{5}$$

where $\Delta_i^{a,b}(t)$ is

$$\Delta_{c,v}^{a,b} = \begin{cases} \Delta_{c,v}^{QD}, & if\ a=b=QD \\ \Delta_{c,v}^{MV}, & if\ a=b=MV \\ (\Delta_{c,v}^{QD} + \Delta_{c,v}^{MV})/2, & if\ a \neq b \end{cases} \tag{6}$$

Note that all CB states have the same correction $\Delta_c^{a,b}$, while all VB states have the same correction $\Delta_v^{a,b}$. Corrected wavefunctions $\varphi_i'(r,t)$ now can be calculated by re-diagonalizing $H'(t)$ in $\{\varphi_i(r,t)\}$ subspace. In Supplementary Information Section III we show values for all the correction terms, along with the comparison among the raw LDA, hybrid functionals, LDA after correction and experimental data.

**VI. Thermalization and nuclear vibrational properties.** QD-$MV^{2+}$ systems have first been thermalized at 300 K (room temperature), using Born–Oppenheimer molecular dynamics (BOMD) with Nosé–Hoover thermostat[32, 33]. During the MD procedure, the mass of the ligand pseudo-hydrogens is set to the oxygen mass (15.999 proton mass) to avoid the potential risk of the unphysical (Cd, Se)-H vibrational modes. After 3 ps thermalization, the QD-$MV^{2+}$ systems have reached the thermal-equilibrium state. The phonon spectrum extracted from the BOMD (see Figure S5 in Supplementary Information) agrees well with previous work[34, 35].

**VII. TDDFT-NAMD, frozen-ion TDDFT and BOMD procedures.** In a real-time TDDFT framework, the electronic wavefunctions are given by the TD extension of the Kohn-Sham single particle equation, which reads

$$H(\{\boldsymbol{R}_\kappa(t)\}, \rho(r,t), t)\varphi_i(r,t) = i\,\partial\varphi_i(r,t)/\partial t, \quad \rho(r,t) = \sum_i |\varphi_i(r,t)|^2, \quad (7)$$

where $\boldsymbol{R}_\kappa(t)$ is the nuclear position for atom $\kappa$, $\varphi_i(r,t)$ is the TD wavefunction, and $\rho(r,t)$ is the electron charge density. The nuclear motion is treated classically following Newton's law using *ab initio* calculated forces,

$$M_\kappa\, d^2\boldsymbol{R}_\kappa(t)/dt^2 = \boldsymbol{F}_\kappa(t), \quad \boldsymbol{F}_\kappa(t) = -\partial E(t)/\partial \boldsymbol{R}_\kappa, \quad (8)$$

where $\boldsymbol{F}_\kappa(t)$ is the *ab initio* force on the $\kappa$-th atom with mass $M_\kappa$, and $E$ is the DFT total energy. These two equations constitutes the so-called Ehrenfest dynamics[36]. Real-time TDDFT calculations are done using the PEtot package[37]. PEtot is a plane wave non-local pseudopotential code designed to tackle large systems (~$10^3$ atoms). Its high-speed real-time TDDFT module has been recently developed to extend the capability to non-adiabatic processes in the sub-picosecond region. In our system, a set of norm-conserving pseudopotentials has been used with a 50 Ry cutoff energy. The local density approximation (LDA) has been used for the exchange-correlation potential $V_{XC}$. We have also tested the generalized gradient approximation (GGA) with van der Waals term, but have only found negligible differences (see Figure S2 and Table S1 in Supplementary Information).

At t=0, in TDDFT-NAMD and frozen-ion TDDFT calculations, one electron is excited from the QD HOMO to the QD LUMO to mimic the photoexcitation. Subsequently (t>0), our high-efficient rt-TDDFT code is applied to the photoexcited states to deliver the TD electron wavefunctions by solving the TD Schrödinger equation with a time step of 0.05 fs for the electronic system, and let the nuclei move following Newton's law using the ab initio calculated forces (for TDDFT-NAMD) or freezing all nuclei (for frozen-ion TDDFT). For BOMD, there is no initial (photo) excitation.

**VIII. Vibrational modes and electron-vibrational coupling calculation.** To find the vibrational frequencies $\omega_\nu$ and eigendisplacements (normal modes) $\xi^\nu_{\kappa\alpha}$ we solve the dynamical equation:

$$\sum_{\kappa\alpha} D_{\kappa\alpha,\kappa'\beta}\xi^\nu_{\kappa\alpha} = \omega_\nu^2 \xi^\nu_{\kappa'\beta}, \quad (9)$$

$$D_{\kappa\alpha,\kappa'\beta} := \frac{1}{\sqrt{M_\kappa M_{\kappa'}}} \frac{\partial^2 E^{BO}}{\partial R_{\kappa\alpha} \partial R_{\kappa'\beta}}, \quad (10)$$

where $\kappa$ and $\kappa'$ are the nucleus indices, $\alpha$ and $\beta$ are the Cartesian coordinate indices, $M_\kappa$ is the nuclear mass of atom $\kappa$, $E^{BO}$ is the BO energy of the system and $\nu$ is the index of phonon branch. Note that $\xi^\nu$ are vectors of 3$N$ components ($N$ being the number of atoms) corresponding to the 3$N$-6 vibrational modes.

We perform the calculation of the vibrational frequencies and normal modes using Quantum Espresso[38]. We use a plane-wave cutoff of 50 Ry; the simulation cell is $30 \times 30 \times 30$ Å$^3$; the exchange correlation functional is BLYP[39, 40] and we use HGH norm-conserving pseudopotentials[41]. The total force on the atoms after relaxation is 7.8×10$^{-5}$ Ry/a.u.

Treating electron-vibrational interaction as a perturbation $H'$ of the ground-state Hamiltonian $H^0$ (being the perturbation linear in the nuclear displacements) we have

$$H' \equiv \sum_{\kappa\alpha} \frac{\partial H^0}{\partial R_{\kappa\alpha}} u_{\kappa\alpha} = \sum_\nu \sum_{ij} g^\nu(i,j)(\hat{a}_\nu + \hat{a}_\nu^\dagger)\hat{c}_j^\dagger \hat{c}_i, \qquad (11)$$

where $u_{\kappa\alpha}$ are the nuclear displacements, $i$ and $j$ are the indices of electronic state, $\hat{a}_\nu$, $\hat{a}_\nu^\dagger$ are the phonon annihilation and creation operators and $\hat{c}_i$, $\hat{c}_i^\dagger$ are the electron annihilation and creation operators; $g^\nu(i,j)$ are the electron-phonon coupling matrix elements, which are given by:

$$g^\nu(i,j) \equiv \sum_{\kappa\alpha} \sqrt{\frac{\hbar}{2M_\kappa \omega_\nu}} \xi_{\kappa\alpha}^\nu \left\langle j \left| \frac{\partial H^0}{\partial R_{\kappa\alpha}} \right| i \right\rangle \qquad (12)$$

and were calculated via finite-difference following reference[18].

**IX. Expansion of the MD trajectories as combination of vibrational modes.** The TD nuclear positions $\{\boldsymbol{R}_\kappa(t)\}$ are obtained from MD calculations, where $\kappa$ is the atom index. The eigendisplacements for the molecule are given as $\{\xi_{\kappa\alpha}^\nu\}$ with eigenfrequencies $\omega_\nu$. Before we do the expansion from $\{\boldsymbol{R}_\kappa(t)\}$ in terms of $\{\xi_{\kappa\alpha}^\nu\}$, we first need to remove the global drifts and rotations from $\{\boldsymbol{R}_\kappa(t)\}$,

$$\bar{\boldsymbol{R}}_\kappa(t) = S(t) \cdot \left(\boldsymbol{R}_\kappa(t) - \boldsymbol{R}_C(t)\right) + \boldsymbol{R}_C(0), \qquad (13)$$

where

$$\boldsymbol{R}_C(t) = \frac{1}{N}\sum_{\kappa=1}^N \boldsymbol{R}_\kappa(t), \qquad \boldsymbol{R}_C(0) = \frac{1}{N}\sum_{\kappa=1}^N \boldsymbol{R}_\kappa(0), \qquad (14)$$

while the 3x3 rotation matrix $S(t)$ is obtained from

$$M(t) = \boldsymbol{R}_C(t) \cdot \left(\boldsymbol{R}_C(0)\right)^T, \qquad M = U\Sigma V^\dagger, \qquad S = VU^\dagger, \qquad (15)$$

where the matrices $U$, $\Sigma$ and $V$ are the singular value decompositions of matrix $M$.

Then the expansion of $\{\bar{R}_\kappa(t)\}$ as $\{\xi^\nu_{\kappa\alpha}\}$ can be written as

$$a_\nu(t) = \sum_{\kappa\alpha} \xi^{\nu\,*}_{\kappa\alpha} \cdot \bar{R}_{\kappa\alpha}(t), \tag{16}$$

where $\alpha$ are the three Cartesian dimensions. Finally, we Fourier transform the coefficients

$$a_\nu(\omega_k) = \sum_{n=0}^{N-1} a_\nu(t) e^{-\frac{i2\pi nk}{N_t}}, \qquad k = 0, 1, \dots, N_t - 1, \tag{17}$$

where $N_t$ is the total number of time steps, $\Delta t$ is the time interval of the MD calculation, and $\omega_k = 2\pi k/N_t \Delta t$.

# References

1. Dexter, D. L., A theory of sensitized luminescence in solids. *The Journal of Chemical Physics* 1953, 21, 836-850.
2. Förster, T., Transfer mechanisms of electronic excitation energy. *Radiation Research Supplement* 1960, 326-339.
3. Grimaldi, G.; Crisp, R. W.; ten Brinck, S.; Zapata, F.; van Ouwendorp, M.; Renaud, N.; Kirkwood, N.; Evers, W. H.; Kinge, S.; Infante, I., Hot-electron transfer in quantum-dot heterojunction films. *Nature communications* 2018, 9, 2310.
4. Marcus, R. A., On the theory of oxidation-reduction reactions involving electron transfer. I. *The Journal of Chemical Physics* 1956, 24, 966-978.
5. Ma, J.; Wang, Z.; Wang, L.-W., Interplay between plasmon and single-particle excitations in a metal nanocluster. *Nature communications* 2015, 6, 10107.
6. Wang, Z.; Li, S.-S.; Wang, L.-W., Efficient real-time time-dependent density functional theory method and its application to a collision of an ion with a 2D material. *Physical review letters* 2015, 114, 063004.
7. Kilina, S.; Velizhanin, K. A.; Ivanov, S.; Prezhdo, O. V.; Tretiak, S., Surface ligands increase photoexcitation relaxation rates in CdSe quantum dots. *ACS nano* 2012, 6, 6515-6524.
8. Falke, S. M.; Rozzi, C. A.; Brida, D.; Maiuri, M.; Amato, M.; Sommer, E.; De Sio, A.; Rubio, A.; Cerullo, G.; Molinari, E., Coherent ultrafast charge transfer in an organic photovoltaic blend. *Science* 2014, 344, 1001-1005.
9. Zhu, H.; Yang, Y.; Hyeon-Deuk, K.; Califano, M.; Song, N.; Wang, Y.; Zhang, W.; Prezhdo, O. V.; Lian, T., Auger-assisted electron transfer from photoexcited semiconductor quantum dots. *Nano letters* 2014, 14, 1263-1269.
10. Long, R.; Prezhdo, O. V., Time-domain ab initio analysis of excitation dynamics in a quantum dot/polymer hybrid: atomistic description rationalizes experiment. *Nano letters* 2015, 15, 4274-4281.
11. Klimov, V. I.; McBranch, D. W., Femtosecond 1 P-to-1 S electron relaxation in strongly confined semiconductor nanocrystals. *Physical Review Letters* 1998, 80, 4028.
12. Harris, C.; Kamat, P. V., Photocatalysis with CdSe nanoparticles in confined media: mapping charge transfer events in the subpicosecond to second timescales. *Acs Nano* 2009, 3, 682-690.
13. Wang, Y.-F.; Wang, H.-Y.; Li, Z.-S.; Zhao, J.; Wang, L.; Chen, Q.-D.; Wang, W.-Q.; Sun, H.-B., Electron extraction dynamics in CdSe and CdSe/CdS/ZnS quantum dots adsorbed with methyl viologen. *The Journal of Physical Chemistry C* 2014, 118, 17240-17246.
14. Wu, K.; Li, Q.; Du, Y.; Chen, Z.; Lian, T., Ultrafast exciton quenching by energy and electron transfer in colloidal CdSe nanosheet–Pt heterostructures. *Chemical Science* 2015, 6, 1049-1054.
15. Hyeon-Deuk, K.; Kim, J.; Prezhdo, O. V., Ab initio analysis of Auger-assisted electron transfer. *The journal of physical chemistry letters* 2015, 6, 244-249.

16. Zeng, P.; Kirkwood, N.; Mulvaney, P.; Boldt, K.; Smith, T. A., Shell effects on hole-coupled electron transfer dynamics from CdSe/CdS quantum dots to methyl viologen. *Nanoscale* 2016, 8, 10380-10387.

17. Ischenko, A. A.; Weber, P. M.; Miller, R. D., Capturing chemistry in action with electrons: realization of atomically resolved reaction dynamics. *Chemical reviews* 2017, 117, 11066-11124.

18. Han, P.; Bester, G., First-principles calculation of the electron-phonon interaction in semiconductor nanoclusters. *Physical Review B* 2012, 85, 235422.

19. Han, P.; Bester, G., Carrier relaxation in colloidal nanocrystals: Bridging large electronic energy gaps by low-energy vibrations. *Physical Review B* 2015, 91, 085305.

20. Cˇapek, R. K.; Lambert, K.; Dorfs, D.; Smet, P. F.; Poelman, D.; Eychmüller, A.; Hens, Z., Synthesis of extremely small CdSe and bright blue luminescent CdSe/ZnS nanoparticles by a prefocused hot-injection approach. *Chemistry of Materials* 2009, 21, 1743-1749.

21. Jasieniak, J.; Smith, L.; Van Embden, J.; Mulvaney, P.; Califano, M., Re-examination of the size-dependent absorption properties of CdSe quantum dots. *The Journal of Physical Chemistry C* 2009, 113, 19468-19474.

22. Kilina, S. V.; Neukirch, A. J.; Habenicht, B. F.; Kilin, D. S.; Prezhdo, O. V., Quantum zeno effect rationalizes the phonon bottleneck in semiconductor quantum dots. *Physical review letters* 2013, 110, 180404.

23. Kuznetsov, A. E.; Beratan, D. N., Structural and electronic properties of bare and capped Cd33Se33 and Cd33Te33 quantum dots. *The Journal of Physical Chemistry C* 2014, 118, 7094-7109.

24. Milliron, D. J.; Hughes, S. M.; Cui, Y.; Manna, L., Colloidal nanocrystal heterostructures with linear and branched topology. *nature* 2004, 430, 190.

25. Hofmann, O. T.; Rangger, G. M.; Zojer, E., Reducing the metal work function beyond pauli pushback: a computational investigation of tetrathiafulvalene and viologen on coinage metal surfaces. *The Journal of Physical Chemistry C* 2008, 112, 20357-20365.

26. Peterson, M. D.; Jensen, S. C.; Weinberg, D. J.; Weiss, E. A., Mechanisms for adsorption of methyl viologen on CdS quantum dots. *ACS nano* 2014, 8, 2826-2837.

27. Morris-Cohen, A. J.; Peterson, M. D.; Frederick, M. T.; Kamm, J. M.; Weiss, E. A., Evidence for a through-space pathway for electron transfer from quantum dots to carboxylate-functionalized viologens. *The Journal of Physical Chemistry Letters* 2012, 3, 2840-2844.

28. Morris-Cohen, A. J.; Frederick, M. T.; Cass, L. C.; Weiss, E. A., Simultaneous determination of the adsorption constant and the photoinduced electron transfer rate for a CdS quantum dot–viologen complex. *Journal of the American Chemical Society* 2011, 133, 10146-10154.

29. Kresse, G.; Furthmüller, J., Efficient iterative schemes for ab initio total-energy calculations using a plane-wave basis set. *Physical review B* 1996, 54, 11169.

30. Bester, G., Electronic excitations in nanostructures: an empirical pseudopotential based approach. *Journal of Physics: Condensed Matter* 2008, 21, 023202.

31. Wei, H.; Luo, J.-W.; Li, S.-S.; Wang, L.-W., Revealing the origin of fast electron transfer in TiO2-based dye-sensitized solar cells. *Journal of the American Chemical Society* 2016, 138, 8165-8174.
32. Nosé, S., A unified formulation of the constant temperature molecular dynamics methods. *The Journal of chemical physics* 1984, 81, 511-519.
33. Hoover, W. G., Canonical dynamics: equilibrium phase-space distributions. *Physical review A* 1985, 31, 1695.
34. Tang, X.; Schneider, T.; Buttry, D. A., A vibrational spectroscopic study of the structure of electroactive self-assembled monolayers of viologen derivatives. *Langmuir* 1994, 10, 2235-2240.
35. Han, P.; Bester, G., Insights about the surface of colloidal nanoclusters from their vibrational and thermodynamic properties. *The Journal of Physical Chemistry C* 2012, 116, 10790-10795.
36. Delos, J. B.; Thorson, W. R., Semiclassical Theory of Inelastic Collisions. II. Momentum-Space Formulation. *Physical Review A* 1972, 6, 720.
37. Wang, L.-W. PEtot package. http://cmsn.lbl.gov/html/PEtot/PEtot.html.
38. Giannozzi, P.; Baroni, S.; Bonini, N.; Calandra, M.; Car, R.; Cavazzoni, C.; Ceresoli, D.; Chiarotti, G. L.; Cococcioni, M.; Dabo, I., QUANTUM ESPRESSO: a modular and open-source software project for quantum simulations of materials. *Journal of physics: Condensed matter* 2009, 21, 395502.
39. Becke, A. D., Density-functional exchange-energy approximation with correct asymptotic behavior. *Physical review A* 1988, 38, 3098.
40. Lee, C.; Yang, W.; Parr, R. G., Development of the Colle-Salvetti correlation-energy formula into a functional of the electron density. *Physical review B* 1988, 37, 785.
41. Hartwigsen, C.; Gœdecker, S.; Hutter, J., Relativistic separable dual-space Gaussian pseudopotentials from H to Rn. *Physical Review B* 1998, 58, 3641.

**Acknowledgements.**

We acknowledge financial support from the German Research Foundation (DFG) via the Cluster of Excellence "The Hamburg Centre for Ultrafast Imaging" (CUI). We thank Horst Weller for fruitful discussions and access to a synthesis lab and Lin-Wang Wang for fruitful discussions. The ab-initio calculations were done at the High-Performance Computing Center of the Hamburg University and the High-Performance Computing Center Stuttgart (HLRS).

**Competing interests.** The authors declare no competing interests.